\documentclass[twocolumn,aps,pra,longbibliography,floatfix]{revtex4-1}
\usepackage[T2A]{fontenc}
\usepackage[utf8]{inputenc}
\usepackage[russian, english]{babel}
\usepackage{graphicx}
\usepackage{nicefrac}
\usepackage{amsmath}
\usepackage{amsfonts}
\usepackage{amssymb}
\usepackage{amsthm}
\usepackage{float}
\usepackage{verbatim}
\usepackage{epsf}
\usepackage{bm}
\usepackage{bbm}
\usepackage{graphicx}
\usepackage{hyperref}
\usepackage{braket}
\usepackage{indentfirst}
\usepackage{siunitx}
\usepackage[table,xcdraw,dvipsnames]{xcolor}
\usepackage{booktabs}

\usepackage{dcolumn}
\usepackage{physics}
\newcommand{\vare}{\varepsilon}
\newcommand{\bfr}{{\bm r}}
\newcommand{\balpha}{\bm{\alpha}}
\newcolumntype{w}[1]{D{.}{.}{#1}}

\begin{document}

\title{
QED calculations of the $\bf E1$ transition amplitude in neon-like iron and nickel
}
\author{M.~G.~Kozlov$^{1,2}$}
\author{V.~A.~Yerokhin$^3$}
\author{M.~Y.~Kaygorodov$^1$}
\author{E.~V.~Tryapitsyna$^1$}

\affiliation{$^1$ Petersburg Nuclear Physics Institute of NRC ``Kurchatov Institute'', Gatchina, Leningrad District 188300, Russia \\
$^2$ Department of Physics, St. Petersburg Electrotechnical University LETI, St. Petersburg 197376, Russia\\
$^3$ 
Peter the Great St. Petersburg Polytechnic University, Polytekhnicheskaya 29, 195251 St. Petersburg, Russia
}

\begin{abstract}

We calculated QED corrections to the $E1$ transition amplitudes in Ne-like iron and nickel. For the $2p \to 3d$ transitions the dominant effect came from the many-electron mixing, or electronic correlations. For the $2p \to 3s$ transitions the correlation and one-electron effects were comparable and tended to compensate each other. Our \textit{ab initio} calculations showed that vertex corrections were negligible for both types of transitions. Other QED corrections were accurately reproduced by including effective QEDMOD operator in the many-electron relativistic configuration interaction calculation.

\end{abstract}

\maketitle

\section{Introduction}

Recent high accuracy  calculations of $E1$ transition amplitudes in neon-like iron and nickel revealed rather large QED corrections for such relatively low-$Z$ systems  \cite{Kuhn2020,KCO22,SKBS24}. This increased the interest to QED corrections to the transition amplitudes in many-electron atoms and ions. 

There are many calculations of the QED corrections to the $E1$ transition amplitudes in highly-charged one- and two-electron ions, see for instance Refs. \cite{SaPaChe04,Indelicato2004,Andreev2009}. For the many-electron neutral atoms, or weakly charged ions, such corrections are usually not included. Neutral Cs is probably the most prominent exception. It was used for the high precision measurement of the parity non-conservation in atoms \cite{WBC97}. Using atomic theory one can accurately determine from this experiment the weak charge of the nucleus and test Standard Model at low energies \cite{SBDJ18}. This is why QED corrections to the $E1$ amplitudes in Cs were calculated several times in different approaches. \citet{SaChe05} calculated QED correction to the $6s_{1/2} \to 6p_{1/2}$ amplitude {\em ab initio} in the one-electron approximation, while calculations \cite{Fairhall2023,TaDe23,Roberts2023} included correlations but used an approximate model QED potential \cite{FlaGin05} and neglected the vertex correction.

Here we analyse in detail QED corrections to the $E1$ amplitudes in neon-like Fe$^{16+}$ and Ni$^{18+}$, explain their surprisingly large size, and discuss implications for other systems.

\section{One-electron approximation}

\textit{Ab initio} QED calculations are presently feasible in the
one-electron approximation, for local screening potentials (not the Dirac-Fock
potential). For this reason, we start with carrying out \textit{ab initio} QED calculations
in the one-electron approximation, compare the \textit{ab initio} results with the approximate QED
treatment, and then proceed to the many-electron calculations.

%
%
\subsection{\textit{Ab initio} QED}

The $E1$ transition amplitude between one-electron states $a$ and $b$
is proportional to the matrix element of the electric dipole
operator in the length gauge,
\begin{align}
\label{eq:zab}
z_{ab} = \bra{a} r_z \ket{b}
= A(j_a,m_a,j_b,m_b) \mathcal{I}(a,b)
\,,
\end{align}
where $r_z$ is the $z$ component of the position vector $\bm{r}$; $A$ is the angular factor, which depends on the angular momenta $j_a,j_b$ and their projections $m_a,m_b$; $\mathcal{I}(a,b)$ is the radial integral, which depends on the radial parts of the upper and lower components $g$ and $f$ of the electronic wavefunction:
\begin{align}
\label{eq:Iab}
\mathcal{I}(a,b) = \int_0^\infty (g_a g_b+f_a f_b) r^3 dr
\,.
\end{align}

The one-loop QED corrections to the
matrix element $z_{ab}$ are induced by the electron self-energy (se)
and vacuum polarization (vp),
\begin{align}
\delta z_{\rm qed} = \delta z_{\rm se} + \delta z_{\rm vp}\,.
\end{align}
General expressions for the QED corrections to the $E1$ transition
amplitude can be found
in Ref.~\cite{SaChe05} (see also Ref.~\cite{shabaev:02:rep}).
The self-energy correction is given by
the sum of the perturbed-orbital (po) and the vertex$+$reducible (vr) parts,
\begin{align}
\delta z_{\rm se} = \delta z_{\rm se, po} + \delta z_{\rm se, vr}\,.
\end{align}
The perturbed-orbital part is given by
\begin{align} \label{qed:po}
\delta z_{\rm se, po} = &\
    \sum_{n\neq a}\bra{a} \Sigma_R(\vare_a) \ket{n}\,
        \frac{z_{nb}}{\vare_a-\vare_n}
    \nonumber \\ &
+    \sum_{n\neq b}
        \frac{z_{an}}{\vare_b-\vare_n}
        \bra{n} \Sigma_R(\vare_b) \ket{b}\,,
\end{align}
where $\Sigma_R(\vare)$ is the renormalized one-loop self-energy operator, $\Sigma_R(\vare) = \Sigma(\vare)-\gamma^0\delta m$,
and $\delta m$ is the one-loop mass counterterm.
The one-loop self-energy operator is defined by its matrix elements with one-electron wave functions
as \cite{yerokhin:99:pra}
\begin{eqnarray} \label{se6}
\bra{a} \Sigma(\vare) \ket{b} = \frac{i}{2\pi} \int_{-\infty}^{\infty}d\omega\,
 \sum_n \frac{\bra{an} I(\omega) \ket{nb}}{\vare-\omega -u\vare_n}\,,
\end{eqnarray}
where $u = 1-i0$, the summation is performed over the complete spectrum of the Dirac equation and the operator $I(\omega)$ describes the
exchange of a virtual photon. In the Feynman gauge, $I(\omega)$ is given by
\begin{eqnarray} \label{se6b}
I(\omega,\bfr_1,\bfr_2) = \alpha \big( 1-\balpha_1\cdot \balpha_2\big)\, \frac{e^{i\sqrt{\omega^2+i0}\,r_{12}}}{r_{12}}\,,
\end{eqnarray}
where $\balpha$ is vector of Dirac matrices and $r_{12} = |\bfr_1-\bfr_2|$.
The vertex$+$reducible contribution is given by
\begin{align}
\delta z_{\rm se, vr} = &\
    \frac{i}{2\pi}\int_{-\infty}^{\infty}d\omega\,
  \nonumber \\ &
    \times
    \Bigg[
        \sum_{n_1n_2}\frac{z_{n_1n_2}\, \bra{an_2} I(\omega) \ket{n_1b}}
            {(\vare_a-\omega-u\vare_{n_1})(\vare_b-\omega-u\vare_{n_2})}
  \nonumber \\ &
  - \frac{z_{ab}}{2}\,\sum_{n}\frac{\bra{an} I(\omega) \ket{na}}
            {(\vare_a-\omega-u\vare_{n})^2}
  \nonumber \\ &
  - \frac{z_{ab}}{2}\,\sum_{n}\frac{\bra{bn} I(\omega) \ket{nb}}
            {(\vare_b-\omega-u\vare_{n})^2}
  \Bigg]\,.
\end{align}
For renormalization purposes, we separate from the vertex$+$reducible contribution 
the contribution of free electron proparagors. This contribution is
denoted as the zero-potential part and is calculated in
the momentum space. The remainder (denoted as the many-potential part) contains one or more
interactions with the binding Coulomb potential in electron propagators. It is
ultraviolet finite and calculated in coordinate space. The separation
into the zero-potential and many-potential parts is 
\begin{align}
\delta z_{\rm se, vr} = \delta z_{\rm se, vr}^{(0)} + \delta z_{\rm se, vr}^{(1+)}\,.
\end{align}

The vacuum-polarization correction is well approximated by its perturbed-orbital part, $\delta z_{\rm vp}\approx \delta z_{\rm vp, po}$, which has the same form as Eq.~(\ref{qed:po}),
\begin{align}
\delta z_{\rm vp, po} = &\
    \sum_{n\neq a}\bra{a} V_{\rm vp} \ket{n}\,
        \frac{z_{nb}}{\vare_a-\vare_n}
    \nonumber \\ &
+    \sum_{n\neq b}
        \frac{z_{an}}{\vare_b-\vare_n}
        \bra{n} V_{\rm vp} \ket{b}\,,
\end{align}
where $V_{\rm vp}$ is the vacuum-polarization potential, see Ref.~\cite{shabaev:02:rep} for details.

\begin{table*}[htb]
\caption{
Individual QED contributions for one-electron $E1$ transition amplitudes in {Fe}$^{16+}$,
in terms of the function $R$ defined by Eq.\ \eqref{res:1}, calculated 
with the Kohn-Sham screening potential.
\label{tab:abinitio}
}
\begin{ruledtabular}
\begin{tabular}{ccw{2.7}w{2.7}w{2.7}w{2.7}w{2.7}w{2.7}}
 \multicolumn{1}{c}{$a$} &
     \multicolumn{1}{c}{$b$} &
         \multicolumn{1}{c}{$R_{\rm se, po}$} &
             \multicolumn{1}{c}{$R_{\rm se, vr}^{(0)}$} &
                 \multicolumn{1}{c}{$R_{\rm se, vr}^{(1+)}$} &
             \multicolumn{1}{c}{$R_{\rm se, vr}$} &
                 \multicolumn{1}{c}{$R_{\rm vp}$} &
                     \multicolumn{1}{c}{$R_{\rm qed}$} 
                      \\
                         \hline\\[-5pt]
 $2p_{1/2}$ & $3s$       &  -0.5446\,(1)  & -1.2899\,(8)  & 1.2943\,(10) & 0.0044\,(13) & 0.0407 & -0.4995\,(13) \\ 
 $2p_{3/2}$ & $3s$       &  -0.4799\,(1)  & -1.2441\,(8)  & 1.2475\,(10) & 0.0034\,(13) & 0.0397 & -0.4368\,(13) \\ 
 $2p_{1/2}$ & $3d_{3/2}$ &  -0.0074       & -0.1604\,(4)  & 0.1609\,(1)  & 0.0005\,(4)  &-0.0002 & -0.0071\,(4) \\ 
 $2p_{3/2}$ & $3d_{3/2}$ &   0.0093       & -0.1602\,(4)  & 0.1596\,(1)  &-0.0006\,(4)  & 0.0000 &  0.0087\,(4) \\ 
 $2p_{3/2}$ & $3d_{5/2}$ &   0.0074\,(1)  & -0.1580\,(4)  & 0.1593\,(2)  & 0.0013\,(4)  & 0.0000 &  0.0086\,(5) \\ 
\end{tabular}
\end{ruledtabular}
\end{table*}

Our numerical calculations of the self-energy and vacuum-polarization
corrections were carried out for 
the Kohn-Sham (KS) local screening potential \cite{1965_KohnW_PR140}. 
For representing the electron propagators we numerically solve the
Dirac equation with the nuclear Coulomb potential and the KS 
screening potential. The
Green’s function of the Dirac equation in a general
(asymptotically Coulomb) potential was computed by the
method described in Appendix of Ref.~\cite{yerokhin:11:fns}.

Numerical results for QED corrections to the $E1$ transition amplitude
are conveniently represented
in terms of the multiplicative function $R_{\rm qed}(Z)$ defined as
\begin{align}\label{res:1}
\delta z_{\rm qed} = z_{ab}\, \frac{\alpha}{\pi}\,R_{\rm qed}(Z)\,.
\end{align}
We thus represent $R_{\rm qed}$ as
\begin{align}\label{res:2}
R_{\rm qed} = R_{\rm se, po} + R_{\rm se, vr}^{(0)}+ R_{\rm se, vr}^{(1+)} + R_{\rm vp}\,.
\end{align}
The QED correction to the radial integral \eqref{eq:Iab} has the same form as Eq.\ \eqref{res:1}, 
\begin{align}\label{res:1a}
\delta \mathcal{I}_{\rm qed}(a,b) = \mathcal{I}(a,b)\, \frac{\alpha}{\pi}\,R_{\rm qed}(Z)\,.
\end{align}

Our numerical results for individual QED contributions for {Fe}$^{16+}$ are presented in Table~\ref{tab:abinitio}. We note a large cancellation between
the zero-potential and many-potential vertex$+$reducible contributions, $R_{\rm se, vr}^{(0)}$ and $R_{\rm se, vr}^{(1+)}$. Separately, these contributions are of the same order as the perturbed-orbital part, but their sum is by about three orders of magnitude smaller. Because of this cancellation, a very careful control of the numerical accuracy is required at intermediate stages of the computation. We also note that the perturbed-orbital contribution
dominates over the vertex$+$reducible part. 

%
%
\subsection{Model QED operator}

Here we compare the results of \textit{ab initio} QED calculations with the approximate
results obtained with an effective QED operator within quantum mechanical approach. There were several such operators suggested in the literature \cite{FlaGin05,TuBe13,STY13,TKSSD16}.
These operators are now broadly used for calculations of many-electron systems including atoms
\cite{GiBe16,GuPaNa24,SkPrMa24}, highly-charged ions \cite{PSSS20,AlDzFl24}, molecules \cite{SkChSh21,Sk21,WiPeUd24} and superheavy elements \cite{ElFrKa15,MaGlSh22}.

Here we use model QED operator formulated in Ref.~\cite{STY13}
and implemented as the QEDMOD package in Refs.~\cite{STY15,shabaev:18:qedmod}. 
This approximate treatment does not include the self-energy vertex and reducible corrections
but should approximately reproduce the perturbed-orbital part of the self-energy and vacuum-polarization corrections,
\begin{align}\label{res:3}
R_{\rm qed}^{\rm mod} \approx R_{\rm se, po} + R_{\rm vp, po}\,.
\end{align}
Fortunately, for the $E1$ transition amplitudes the vertex and reducible corrections are typically small and the QEDMOD operator provides a reasonable approximation for the total QED correction.

\begin{table}[htbp]
\caption{QED corrections to one-electron radial integrals $\mathcal{I}(a,b)$ for Fe$^{16+}$. Comparison of the \textit{ab initio} calculation with QEDMOD calculations for DF and KS potentials. 
}
\label{tab:Fe_rad_int}
{\begin{tabular}{cc| w{1.6}w{2.3}| w{1.6}w{2.3}| w{2.8}}
\hline
\hline
&&\multicolumn{4}{c|}{QEDMOD}
&\multicolumn{1}{c}{\textit{ab initio}}
\\
\cline{3-7}
\multicolumn{2}{c|}{Transition}&\multicolumn{2}{c|}{DF potential}
&\multicolumn{3}{c}{KS potential}
\\
\cline{3-7}
$a$ & $b$
&\multicolumn{1}{c}{$\mathcal{I}(a,b)$}
&\multicolumn{1}{c|}{$R_\mathrm{qed}^\mathrm{mod}$}
&\multicolumn{1}{c}{$\mathcal{I}(a,b)$}
&\multicolumn{1}{c|}{$R_\mathrm{qed}^\mathrm{mod}$}
&\multicolumn{1}{c}{$R_\mathrm{qed}$}
\\
\hline
 $2p_{1/2}$ & $3s_{1/2}$ & 0.05905 & -0.488 &  0.05749 & -0.494 &  -0.4995\,(13) \\
 $2p_{3/2}$ & $ 3s_{1/2}$ & 0.06187 & -0.422 &  0.06048 & -0.425 &  -0.4368\,(13) \\
 $2p_{1/2}$ & $ 3d_{3/2}$ & 0.18750 & -0.016 &  0.18993 & -0.005 &  -0.0071\,(4) \\
 $2p_{3/2}$ & $ 3d_{3/2}$ & 0.19067 &  0.003 &  0.19329 &  0.013 &   0.0087\,(4) \\
 $2p_{3/2}$ & $ 3d_{5/2}$ & 0.19076 &  0.002 &  0.19330 &  0.011 &   0.0086\,(5) \\
\hline\hline 
\end{tabular}
}
\end{table}
In Table \ref{tab:Fe_rad_int} we present QED corrections to the $E1$ radial integrals between valence electron states of Fe$^{16+}$. \textit{Ab initio} QED corrections were computed for the KS potential. The approximate calculations with the QEDMOD effective operator were performed for both the KS and the non-local Dirac-Fock (DF) potentials. Looking at this table we can make two observations.

First, for the $s$-$p$ transitions the results of the \textit{ab initio} and QEDMOD calculations differ by 2-3\%. Calculations with the local KS and non-local Dirac-Fock potentials differ by approximately 1\%. We conclude that here we have very good agreement and the difference between \textit{ab initio} QED calculation and QEDMOD calculation is comparable to the difference between local and non-local mean-field potentials. The vertex contribution is smaller than the differences mentioned above. Therefore, at the present level of the theory, the vertex corrections can be neglected.
   
Second, QED corrections to the $p$-$d$ radial integrals are more than an order of magnitude smaller than for the $s$-$p$ transitions. The difference between local and non-local potentials is huge. This means that the indirect effect from the adjustment of the self-consistent field is larger than the direct QED contribution. The difference between \textit{ab initio} and QEDMOD calculations is smaller, but also quite significant, about 30-50\%. Compared to these differences, the vertex corrections are completely negligible. 
\section{Many-electron calculations}

In this section we describe our results for the QED corrections to the energies and transition rates in the ten-electron ions Fe$^{16+}$ and  Ni$^{18+}$. The many-electron $E1$ transition amplitude between the states with the total angular momentum $J$ and parity $p$ can be described by the reduced matrix element. In the L-gauge it can be written in the following form:
\begin{multline}
    \label{eq:reducedME}
    E1(J'p',Jp) \equiv
    \langle J'p'||E1||Jp\rangle 
    =\sum_{nlj,n'l'j'} \rho_{n'l'j',nlj}^{1} 
    \\
    \times (-1)^{j+l_\mathrm{max}-1/2}
    \sqrt{(2j'+1)(2j+1)l_\mathrm{max}}
    \\
    \times \left\{
    \begin{array}{ccc}
         l'& j'& 1/2  \\
         j & l & 1 
    \end{array}
    \right\}
    \mathcal{I}(n'l'j',nlj)\,,
\end{multline}
where $\rho_{n'l'j',nlj}^{1}$ is the reduced transition matrix (RTM) of the tensor rank 1, $\mathcal{I}(n'l'j',nlj)$ is the respective radial integral \eqref{eq:Iab}, and $l_\mathrm{max} =\max(l,l')$. RTM of the rank $L$ is defined as follows:
\begin{align}
&\rho^L_{n'l'j',nlj} = (-1)^{J'-M'}
\left(\!\!\!\begin{array}{rcc}
 J' & L & J \\
-M' & q & M \\
\end{array}\!\!\right)^{-1}
\nonumber \\
&\times\sum_{mm'} (-1)^{j'-m'} \left(\!\!\! \begin{array}{rcc}
 j' & L & j \\
-m' & q & m \\
\end{array} \!\!\right) \rho_{n'l'j'm',nljm} ,
\label{rhoL}
\end{align}
where
\begin{align}\label{tm1a}
 &\rho_{n'l'j'm',nljm} = \langle J'M'| a^\dagger_{n'l'j'm'}
 a_{nljm}|JM\rangle\,,
\end{align}
and $a^\dagger_{n'l'j'm'}$ and $a_{nljm}$ are creation and annihilation operators respectively.

\begin{table*} [htb]
\caption{Transition energies in cm$^{-1}$ from the ground state $^1$S$_0(2s^22p^6)$ for Fe$^{16+}$. Calculation were done using configuration interaction (CI) method for Coulomb-Breit (CB) Hamiltonian either with the Dirac-Fock (DF) orbitals, or with the Kohn-Sham (KS) orbitals. QED corrections $\delta E_\mathrm{qed}^\mathrm{mod}$ are calculated using effective operator QEDMOD.
}
\label{tab:Fe_energies}
{\begin{tabular}{lc rrr rrrr} 
\hline
\hline
&
&\multicolumn{1}{c}{$E_\mathrm{NIST}$}
&\multicolumn{2}{c}{DF potential}
&\multicolumn{2}{c}{KS potential}
&\multicolumn{2}{c}{\citet{Kuhn2020}}
\\
&
&\multicolumn{1}{c}{Ref.\ \cite{NIST}}
&\multicolumn{1}{c}{$E_\mathrm{CB}$}
&\multicolumn{1}{c}{$\delta E_\mathrm{qed}^\mathrm{mod}$}
&\multicolumn{1}{c}{$E_\mathrm{CB}$}
&\multicolumn{1}{c}{$\delta E_\mathrm{qed}^\mathrm{mod}$}
&\multicolumn{1}{c}{$E_\mathrm{CB}$}
&\multicolumn{1}{c}{$\delta E_\mathrm{qed}^\mathrm{mod}$}
\\
\hline
$2s^22p^53p$ & $^3$S$_1$                 & 6093450 & 6090245 &   63  & 6092954 &   61 & 6092092 &   67 \\
$2s^22p^53p$ & $^3$D$_2$                 & 6121690 & 6119014 &   49  & 6121709 &   47 & 6119799 &   43 \\
$2s^22p^53p$ & $^3$D$_3$                 & 6134730 & 6131734 &  101  & 6134428 &   98 & 6132770 &   94 \\
$2s^22p^53p$ & $^1$P$_1$                 & 6143850 & 6141060 &   87  & 6143748 &   84 & 6141943 &   82 \\[4pt]
$2s^22p^53s$ & $(\tfrac32,\tfrac12)_1^o$ & 5864770 & 5862854 &  803  & 5866066 &  801 & 5861842 &  784 \\[2pt]
$2s^22p^53s$ & $(\tfrac12,\tfrac12)_1^o$ & 5960870 & 5958702 & 1055  & 5961947 & 1053 & 5957841 & 1042 \\[2pt]
$2s^22p^53d$ & $^3$P$_1^o$               & 6471800 & 6466471 &   85  & 6468547 &   83 & 6469630 &   87 \\
$2s^22p^53d$ & $^3$D$_1^o$               & 6552200 & 6548612 &  143  & 6550662 &  141 & 6550895 &  134 \\
$2s^22p^53d$ & $^1$P$_1^o$               & 6660000 & 6658447 &  287  & 6660540 &  285 & 6659174 &  288 \\
\hline\hline 
\end{tabular}}
\end{table*}

Note that QED corrections to the many-electron transition amplitudes depend not only on the one-electron radial integrals, but also on the mixing coefficients of the many-electron states. The latter determine the values of the RTM amplitudes in Eq.\ \eqref{eq:reducedME}. 

Calculations in this work were performed by the configuration interaction (CI) method implemented by the computer package CI-MBPT \cite{KPST15}. 
Calculations for both ions were done in the same manner. We used the basis set $[17spdfg]$, which included 4-component orbitals with principle quantum numbers $n\le 17$ for 9 relativistic partial waves from $s_{1/2}$ to $g_{9/2}$. This basis set was formed using the method described in Refs.\ \cite{KoDeKa24,KozTup19}. CI space included all single excitations and most important double and triple excitations from the reference configurations. The sizes of the CI space for both parities were about $10^6$ determinants. 

\subsection{CI calculation of neon-like Fe}

For Fe$^{16+}$ we performed calculations for the DF and KS basis sets. QED corrections were included using QEDMOD potential. In Table \ref{tab:Fe_energies} we give transition energies from the ground state to several low-lying levels of both parities. We see that both calculations are in a good agreement with the experiment, the relative differences in transition energies are about 0.05\%, or less. At the same time, QED corrections for these two calculations differ by up to 5\%. 
This difference can be explained by the core relaxation, which for the KS potential is included only via configurational mixing, while for the DF potential a significant part of this effect is already included in the initial approximation, when the basis set is formed. The last two columns of the table present results of the calculation with a significantly larger CI space in Ref.\ \cite{Kuhn2020}, which used parallelized version \cite{Cheung2021} of the CI code \cite{KPST15}. All three calculations are in good agreement with each other. 

\begin{table} [htb]
\caption{Reduced matrix elements (a.u.) for $E1$ transitions from the ground state $^1$S$_0$ in Fe$^{16+}$. QED corrections are calculated using effective operator QEDMOD.}
\label{tab:Fe_reduced_ME}
\begin{tabular}{cc|rr|rr}
\hline
\hline
\multicolumn{1}{c}{final}
&\multicolumn{1}{c|}{nominal 1e}
&\multicolumn{2}{c|}{DF potential}
&\multicolumn{2}{c}{KS potential}
\\
\multicolumn{1}{c}{state}
&\multicolumn{1}{c|}{transition}
&\multicolumn{1}{c}{$E1_0$}
&\multicolumn{1}{c|}{$\delta E1_\mathrm{qed}^\mathrm{mod}$}
&\multicolumn{1}{c}{$E1_0$}
&\multicolumn{1}{c}{$\delta E1_\mathrm{qed}^\mathrm{mod}$}
\\
\hline
$(\tfrac32,\tfrac12)_1^o$ & $2p\to 3s$     & 0.08374 &$ -0.00001 $&  0.08360 &$ -0.00001 $ \\
$(\tfrac12,\tfrac12)_1^o$ & $2p\to 3s$     & 0.07603 &$ -0.00006 $&  0.07592 &$ -0.00006 $ \\
$^3$P$_1^o$               & $2p\to 3d$     & 0.02231 &$  0.00002 $&  0.02228 &$  0.00002 $ \\
$^3$D$_1^o$               & $2p\to 3d$     & 0.17506 &$  0.00032 $&  0.17488 &$  0.00032 $ \\
\multicolumn{2}{c|}{Ref.\ \cite{Kuhn2020}} & 0.17891 &$  0.00030 $\\ 
$^1$P$_1^o$               & $2p\to 3d$     & 0.33675 &$ -0.00013 $&  0.33644 &$ -0.00012 $ \\
\multicolumn{2}{c|}{Ref.\ \cite{Kuhn2020}} & 0.33515 &$ -0.00017 $\\ 
\hline\hline 
\end{tabular}
\end{table}

Results of the CI calculation of the many-electron $E1$ reduced matrix elements are presented in Table \ref{tab:Fe_reduced_ME}. First, we see that the values calculated with the local and non-local potentials are very close, including the ones for QED corrections. 
The reduced matrix elements for the last two transitions were also calculated in Ref.\ \cite{Kuhn2020} using much larger CI space and the same QEDMOD operator. We see that present calculation slightly underestimate transition amplitude to the $^3$D$_1^o$ state; otherwise our results are in good agreement with \citet{Kuhn2020}. We conclude that QED corrections calculated with QEDMOD potential for Fe$^{16+}$ are numerically stable.

When we compare QED corrections for many-electron reduced matrix elements from Table \ref{tab:Fe_reduced_ME} and for the one-electron radial integrals from Table \ref{tab:Fe_rad_int}, we see a striking difference. The first two transitions in Table \ref{tab:Fe_reduced_ME} are nominally $2p\to 3s$ transitions. Thus, according to Table \ref{tab:Fe_rad_int}, we expect QED corrections on the order of 0.1\%. The last three transitions are, on the contrary, $2p\to 3d$ transitions and QED corrections are supposed to be few orders of magnitude smaller. 

However, we see the opposite results of the many-electron calculations. The smallest relative QED correction is for the first $^1\mathrm{S}_0 \to {}^3\mathrm{D}_1^o$ ($2p\to 3s$) transition, about $-0.01\%$, and the largest correction, about $+0.2\%$, is for the transition $^1S_0\to (\tfrac32,\tfrac12)^o_1$ ($2p\to 3d$). The reason for this is the following: in Eq.\ \eqref{eq:reducedME} not only the radial integral $\mathcal{I}(n'l'j',nlj)$, but also the amplitude $\rho_{n'l'j',nlj}^{1}$ depends on the QED corrections. In many cases the latter dependence turns out to be dominant. For example, the largest contributions to the $^1\mathrm{S}_0 \to {}^3\mathrm{D}_1^o$ transition are:
\begin{multline}
    \label{eq:RME3D1}
    E1\left(^1\mathrm{S}_0 \to {}^3\mathrm{D}_1^o\right) 
    \approx
    \sqrt{\frac43}\rho^1_{3d_{3/2},2p_{1/2}} \mathcal{I}(3d_{3/2},2p_{1/2})
    \\
    -\frac{2}{\sqrt{15}}\rho^1_{3d_{3/2},2p_{3/2}} \mathcal{I}(3d_{3/2},2p_{3/2})
    \\
    +2\sqrt{\frac35}\rho^1_{3d_{5/2},2p_{3/2}} \mathcal{I}(3d_{5/2},2p_{3/2})\,,
\end{multline}
where the coefficients follow from Eq. \eqref{eq:reducedME}. QED corrections to all three radial integrals here are very small, whereas corrections to the RTM amplitudes are much larger. Specifically, the calculated RTM amplitudes have the following values with and without QED corrections, respectively:
\begin{subequations}
\label{eq:tm_values}
\begin{align}
                              & \quad\mathrm{no~QED} &\mathrm{with~QED}  \nonumber \\
 \rho^1_{3d_{3/2},2p_{1/2}} = & -0.43506\,,             & -0.43431\,, \\
 \rho^1_{3d_{3/2},2p_{3/2}} = & -0.50076\,,             & -0.50066\,, \\
 \rho^1_{3d_{5/2},2p_{3/2}} = &\quad 0.74227\,,         & 0.74278\,. 
\end{align}
\end{subequations}

The changes in matrix elements of the RTM account for the main part of the QED correction to the amplitude of this transition. If we substitute values from Table \ref{tab:Fe_rad_int} and Eqs.\ \eqref{eq:tm_values} into Eq.\ \eqref{eq:RME3D1} we get 
\begin{align}
\label{eq:p-d_approx}
  E1_0\left(^1\mathrm{S}_0 \to {}^3\mathrm{D}_1^o\right) &\approx 0.17447 \quad \delta E1_\mathrm{qed}^\mathrm{mod} \approx 0.00031\,, 
\end{align}
in a very good agreement with the values from Table \ref{tab:Fe_reduced_ME}. If we calculate $\delta E1_\mathrm{qed}^\mathrm{mod}$ neglecting QED corrections to the radial integrals, we get $\delta E1_\mathrm{qed}^\mathrm{mod}=0.00030\,$. Clearly, QED corrections to the RTM amplitudes dominate here.  

Now, let us look at the transition $^1S_0\to (\tfrac32,\tfrac12)^o_1$, which is nominally $2p\to 3s$ transition. Two main contributions here are:
\begin{multline}
    \label{eq:RME3/2,1/2}
    E1\left(^1\mathrm{S}_0 \to (\tfrac32,\tfrac12)^o_1\right) 
    \approx
    -\sqrt{\frac23}\rho^1_{3s_{1/2},2p_{1/2}} \mathcal{I}(3s_{1/2},2p_{1/2})
    \\
    -\frac{2}{\sqrt{3}}\rho^1_{3s_{1/2},2p_{3/2}} \mathcal{I}(3s_{1/2},2p_{3/2})\,.
\end{multline}
According to Table \ref{tab:Fe_rad_int}, QED corrections to these radial integrals are larger and CI calculation gives following values of the RTM amplitudes:
\begin{subequations}
\label{eq:tm_values_2p-3s}
\begin{align}
                              & \quad\mathrm{no~QED} &\mathrm{with~QED}  \nonumber \\
 \rho^1_{3s_{1/2},2p_{1/2}} = & \quad\, 0.12438            &  0.12402\,, \\
 \rho^1_{3s_{1/2},2p_{3/2}} = & -0.98782            & -0.98786\,.
\end{align}
\end{subequations}
Substituting values from Table \ref{tab:Fe_rad_int} and Eqs.\ \eqref{eq:tm_values_2p-3s} into Eq.\ \eqref{eq:RME3/2,1/2}, we get 
\begin{align}
\label{eq:p-s_approx}
  E1_0\left(^1\mathrm{S}_0 \to (\tfrac32,\tfrac12)^o_1\right) &\approx 0.06458 
  \quad \delta E1_\mathrm{qed}^\mathrm{mod} \approx 0.00002\,. 
\end{align}
Here the agreement with the value from Table \ref{tab:Fe_reduced_ME} is not as good as in Eq.\ \eqref{eq:p-d_approx}, but nevertheless, these two amplitudes account for more than three quarters of the final answer.
If we calculate $\delta E1_\mathrm{qed}^\mathrm{mod}$ using Eq.\ \eqref{eq:RME3/2,1/2} and neglecting QED corrections to the radial integrals, we get $\delta E1_\mathrm{qed}^\mathrm{mod}=-0.00004\,$. We see that for this amplitude corrections to the radial integrals and to the RTM amplitudes tend to cancel each other. 

We conclude that correlations drastically affect QED corrections to the $E1$ transition amplitudes in Fe$^{16+}$. This is particularly interesting because Fe$^{16+}$ is a highly charged ion with very strong central field and relatively weak configuration mixing. As we see from the above analysis, QED corrections primarily change the fine-structure coupling. For the $p-s$ transitions this effect is comparable to the corrections to radial integrals and dominates for the $p-d$ transitions.

\begin{table} [tbh]
\caption{QED corrections to the $g$ factors of the odd levels from Table \ref{tab:Fe_energies}. CI calculations were done for the CB Hamiltonian and the DF orbitals. QED corrections $\delta g_\mathrm{qed}^\mathrm{mod}$ are calculated using effective operator QEDMOD.}
\label{tab:Fe_g-factors}
{\begin{tabular}{lc crr} 
\hline
\hline
&
&\multicolumn{1}{c}{$E$ (cm$^{-1}$) \cite{NIST}}
&\multicolumn{1}{c}{$g_\mathrm{CB}$}
&\multicolumn{1}{c}{$\delta g_\mathrm{qed}^\mathrm{mod}$}
\\
\hline
$2s^22p^53s$ & $(\tfrac32,\tfrac12)_1^o$ & 5864770 & 1.22364 &$ -0.00018 $\\[2pt]
$2s^22p^53s$ & $(\tfrac12,\tfrac12)_1^o$ & 5960870 & 1.26956 &$  0.00018 $\\[2pt]
$2s^22p^53d$ & $^3$P$_1^o$               & 6471800 & 1.42045 &$ -0.00015 $\\
$2s^22p^53d$ & $^3$D$_1^o$               & 6552200 & 0.65259 &$  0.00040 $\\
$2s^22p^53d$ & $^1$P$_1^o$               & 6660000 & 0.92539 &$ -0.00025 $\\
\hline\hline 
\end{tabular}}
\end{table}

Excited levels of Fe$^{16+}$ belong to the intermediate coupling scheme \cite{Sob79}, where the levels of the $2\bar{p}3d$ configuration are closer to the $LS$ coupling limit (here $2\bar{p}$ stands for the hole in the $2p$ shell). On the other hand, the levels of the $2\bar{p}3s$ configuration are closer to the $jj$ coupling limit. In the first case the residual two-electron Coulomb interaction dominates over the spin-orbit interaction, while in the second case situation is the opposite. QED corrections affect this balance. In the $LS$ coupling scheme there is strong suppression of the transitions between singlet and triplet states, while in the $jj$ coupling there is no such selection rule. Instead, there is suppression of the transitions with $\Delta j>1$. This is why $E1$ transition amplitudes are very sensitive to the coupling scheme. 

Coupling scheme also determines $g$ factors of the electronic states. 
Table \ref{tab:Fe_g-factors} lists $g$ factors of the odd levels with $J=1$, which are linked to the ground state by the $E1$ transitions. We see that the relative QED corrections to the $g$ factors vary from $-0.03\%$ to $+0.06\%$, which is somewhat smaller, but comparable with the QED corrections to the corresponding $E1$ amplitudes. 

\subsection{Using \textit{ab initio} QED corrections in many-electron calculations}

\begin{table*} [htb]
\caption{Reduced matrix elements (a.u.) for $E1$ transitions from the ground state $^1$S$_0$ in Fe$^{16+}$. Corrections $\delta E1_\mathrm{qed}$ are calculated using \textit{ab initio} results for the dominant one-electron transitions and QEDMOD results for subdominant ones; configurational mixing is calculated using QEDMOD. For comparison, the column $\delta E1_\mathrm{qed}^\mathrm{mod}$ repeats results of the pure QEDMOD calculation from Table \ref{tab:Fe_reduced_ME} with more digits. $\delta E1_\mathrm{qed, CI}$ correction includes only configurational mixing. The last column gives contribution of the vertex correction.
}
\label{tab:Fe_reduced_ME_cor}
{\begin{tabular}{cc|rrrrr}
\hline
\hline
\multicolumn{1}{c}{final}
&\multicolumn{1}{c|}{nominal 1e}
&\multicolumn{1}{c}{$E1_0$}
&\multicolumn{1}{c}{$\delta E1_\mathrm{qed}$}
&\multicolumn{1}{c}{$\delta E1_\mathrm{qed}^\mathrm{mod}$}
&\multicolumn{1}{c}{$\delta E1_\mathrm{qed,CI}$}
&\multicolumn{1}{c}{$\delta E1_\mathrm{se,vr}$}
\\
\multicolumn{1}{c}{state}
&\multicolumn{1}{c|}{transition}
\\
\hline
$(\tfrac32,\tfrac12)_1^o$ & $2p\to 3s$ & 0.083738 &$ -0.000011 $&$ -0.000009 $&$  0.000052 $& 0.000001  \\
$(\tfrac12,\tfrac12)_1^o$ & $2p\to 3s$ & 0.076032 &$ -0.000060 $&$ -0.000058 $&$  0.000003 $& 0.000001  \\
$^3$P$_1^o$               & $2p\to 3d$ & 0.022310 &$  0.000019 $&$  0.000018 $&$  0.000020 $& 0.000000  \\
$^3$D$_1^o$               & $2p\to 3d$ & 0.175059 &$  0.000326 $&$  0.000324 $&$  0.000316 $& 0.000000  \\
$^1$P$_1^o$               & $2p\to 3d$ & 0.336746 &$ -0.000119 $&$ -0.000126 $&$ -0.000127 $& 0.000001  \\
\hline\hline 
\end{tabular}}
\end{table*}

In all many-electron $E1$ transitions calculated here the respective nominal one-electron transitions are either $2p \to 3s$, or $2p \to 3d$. For these transitions we made \textit{ab initio} calculations of the QED corrections to one-electron radial integrals, see Table \ref{tab:abinitio}. Thus, in the many-electron expression \eqref{eq:reducedME} we can use the \textit{ab initio} values for the dominant one-electron radial integrals and the QEDMOD values for other radial integrals. In this way we obtain our final and most accurate estimates of the QED corrections to the $E1$ transition amplitudes in Fe$^{16+}$. They are given in the column $\delta E1_\mathrm{qed}$ in Table \ref{tab:Fe_reduced_ME_cor}. For comparison, in other columns of this table we give results of the pure QEDMOD calculation $\delta E1_\mathrm{qed}^\mathrm{mod}$, the calculation with including QEDMOD only into
the configurational mixing $\delta E1_\mathrm{qed,CI}$ (i.e., QEDMOD corrections to the RTM matrix and no QED in the radial integrals), and separately the vertex corrections $\delta E1_\mathrm{se,vr}$. Once again we see that CI mixing is dominant for the $2p\to 3d$ transitions; for the $2p\to 3s$ transitions it is of the same order of magnitude as the QED corrections to the one-electron $E1$ transition amplitudes and has the opposite sign. The vertex corrections are very small in both cases. 

Summing up, we observe that the difference between the $E1$ transition amplitudes calculated with only QEDMOD and with inclusion of \textit{ab initio} QED results is very small, a few parts in $10^{-5}$ relative to $E1_0$. We conclude that despite the fact that the QEDMOD treatment does not include vertex corrections for transition amplitudes, this omission causes only small errors and that in this case QEDMOD is as accurate for the $E1$ amplitudes as for the energies. We recall that a 10\% accuracy for QED corrections calculated with the QEDMOD operator was assumed for the energies of many-electron systems in Refs.~\cite{STY15,TKSSD16}.

\subsection{CI calculation of neon-like Ni}

\begin{table} [htb]
\caption{Reduced matrix elements (a.u.) for $E1$ transitions from the ground state $^1$S$_0$ in Ni$^{18+}$. Corrections $\delta E1_\mathrm{qed}^\mathrm{mod}$ are calculated using QEDMOD operator.}
\label{tab:Ni_reduced_ME}
{\begin{tabular}{cc| rr}
\hline
\hline
\multicolumn{1}{c}{final}
&\multicolumn{1}{c|}{nominal 1e}
&\multicolumn{1}{c}{$E1_0$}
&\multicolumn{1}{c}{$\delta E1_\mathrm{qed}^\mathrm{mod}$}
\\
\multicolumn{1}{c}{state}
&\multicolumn{1}{c|}{transition}
\\
\hline
$(\tfrac32,\tfrac12)_1^o$ & $2p\to 3s$   & 0.077147 &$ -0.000019 $  \\
$(\tfrac12,\tfrac12)_1^o$ & $2p\to 3s$   & 0.066417 &$ -0.000047 $  \\
$^3$P$_1^o$               & $2p\to 3d$   & 0.020621 &$  0.000011 $  \\
$^3$D$_1^o$               & $2p\to 3d$   & 0.182415 &$  0.000280 $  \\
\multicolumn{2}{c|}{Ref.\ \cite{SKBS24}} & 0.185850 &$  0.000280 $\\ 
$^1$P$_1^o$               & $2p\to 3d$   & 0.301420 &$ -0.000122 $  \\
\multicolumn{2}{c|}{Ref.\ \cite{SKBS24}} & 0.300050 &$ -0.000120 $\\ 
\hline\hline 
\end{tabular}}
\end{table}

CI calculation for Ni$^{18+}$ was carried out with the basis set $[17spdfg]$ formed using DF potential. The sizes of the CI spaces for two parities were the same, as in the case of iron. Results are presented in Table \ref{tab:Ni_reduced_ME}. We see again, that the largest QED corrections are for the last two $2p \to 3d$ transitions, while corrections to the $2p\to 3s$ transitions are almost an order of magnitude smaller.

Ni$^{18+}$ has similar level structure as Fe$^{16+}$, but larger nuclear and ionic charges. Thus, one can expect larger relative size of the one-electron QED corrections. Comparison of the results for two ions shows that relative QED corrections for nickel are larger only for the transitions from the ground state to states $(\tfrac32,\tfrac12)_1^o$ and $^1$P$_1^o$. On the contrary, the relative size of the QED transition to the level $^3$D$_1^o$ in nickel is 22\% smaller than in iron. Such behaviour can be easily explained if the most important effect is caused by the transition from $LS$ to $jj$ coupling scheme: this effect is at first linear in QED correction, but then flattens down as we approach the $jj$ limit. The last two transitions from Table \ref{tab:Ni_reduced_ME} were recently calculated in Ref.\ \cite{SKBS24}. QED corrections to these transitions found here are in a very good agreement with the results obtained there.

\section{Conclusions}

At present, purely \textit{ab initio} many-electron QED calculations are impossible for complex many-electron systems. Here we show that the effective QED operator QEDMOD \cite{STY13} accurately includes most important QED contributions into correlated calculation. Vertex corrections, neglected in this approach, turn out to be very small, both for the $p\to s$ and $p\to d$ transitions considered here, at least for the neon-like iron and nickel. We estimate the accuracy of the QEDMOD operator for the QED corrections to the $E1$ transition amplitudes to be about 10\%, same as assumed previously \cite{STY15,TKSSD16} for energies. Our present study confirms high accuracy of the recent calculations of these systems in Refs.\ \cite{Kuhn2020,KCO22,SKBS24}.

QED corrections to the many-electron $E1$ transition amplitudes may be enhanced for the systems with intermediate coupling, where they affect the selection rules implied by a coupling scheme. In the $LS$ limit, the singlet to triplet transitions, like $^1$S$_0\to{}^3$P$_1^o$, are forbidden, while in the $jj$ limit the transitions with $\Delta j>1$, like $p_{1/2}\to d_{5/2}$, are forbidden. Transition from the one coupling to another is the main reason for the large QED corrections to the $2p-3d$ transitions in neon-like iron and nickel.

\begin{acknowledgments}
This work was supported by the Russian Science Foundation grant \#23-22-00079.
\end{acknowledgments}
\bibliography{./qed_fe}

\begin{thebibliography}{38}%
\makeatletter
\providecommand \@ifxundefined [1]{%
 \@ifx{#1\undefined}
}%
\providecommand \@ifnum [1]{%
 \ifnum #1\expandafter \@firstoftwo
 \else \expandafter \@secondoftwo
 \fi
}%
\providecommand \@ifx [1]{%
 \ifx #1\expandafter \@firstoftwo
 \else \expandafter \@secondoftwo
 \fi
}%
\providecommand \natexlab [1]{#1}%
\providecommand \enquote  [1]{``#1''}%
\providecommand \bibnamefont  [1]{#1}%
\providecommand \bibfnamefont [1]{#1}%
\providecommand \citenamefont [1]{#1}%
\providecommand \href@noop [0]{\@secondoftwo}%
\providecommand \href [0]{\begingroup \@sanitize@url \@href}%
\providecommand \@href[1]{\@@startlink{#1}\@@href}%
\providecommand \@@href[1]{\endgroup#1\@@endlink}%
\providecommand \@sanitize@url [0]{\catcode `\\12\catcode `\$12\catcode `\&12\catcode `\#12\catcode `\^12\catcode `\_12\catcode `\%12\relax}%
\providecommand \@@startlink[1]{}%
\providecommand \@@endlink[0]{}%
\providecommand \url  [0]{\begingroup\@sanitize@url \@url }%
\providecommand \@url [1]{\endgroup\@href {#1}{\urlprefix }}%
\providecommand \urlprefix  [0]{URL }%
\providecommand \Eprint [0]{\href }%
\providecommand \doibase [0]{http://dx.doi.org/}%
\providecommand \selectlanguage [0]{\@gobble}%
\providecommand \bibinfo  [0]{\@secondoftwo}%
\providecommand \bibfield  [0]{\@secondoftwo}%
\providecommand \translation [1]{[#1]}%
\providecommand \BibitemOpen [0]{}%
\providecommand \bibitemStop [0]{}%
\providecommand \bibitemNoStop [0]{.\EOS\space}%
\providecommand \EOS [0]{\spacefactor3000\relax}%
\providecommand \BibitemShut  [1]{\csname bibitem#1\endcsname}%
\let\auto@bib@innerbib\@empty
\bibitem [{\citenamefont {K{\"{u}}hn}\ \emph {et~al.}(2020)\citenamefont {K{\"{u}}hn}, \citenamefont {Shah}, \citenamefont {L{\'{o}}pez-Urrutia}, \citenamefont {Fujii}, \citenamefont {Steinbr{\"{u}}gge}, \citenamefont {Stierhof}, \citenamefont {Togawa}, \citenamefont {Harman}, \citenamefont {Oreshkina}, \citenamefont {Cheung}, \citenamefont {Kozlov}, \citenamefont {Porsev}, \citenamefont {Safronova}, \citenamefont {Berengut}, \citenamefont {Rosner}, \citenamefont {Bissinger}, \citenamefont {Ballhausen}, \citenamefont {Hell}, \citenamefont {Park}, \citenamefont {Chung}, \citenamefont {Hoesch}, \citenamefont {Seltmann}, \citenamefont {Surzhykov}, \citenamefont {Yerokhin}, \citenamefont {Wilms}, \citenamefont {Porter}, \citenamefont {St{\"{o}}hlker}, \citenamefont {Keitel}, \citenamefont {Pfeifer}, \citenamefont {Brown}, \citenamefont {Leutenegger},\ and\ \citenamefont {Bernitt}}]{Kuhn2020}%
  \BibitemOpen
  \bibfield  {author} {\bibinfo {author} {\bibfnamefont {Steffen}\ \bibnamefont {K{\"{u}}hn}}, \bibinfo {author} {\bibfnamefont {Chintan}\ \bibnamefont {Shah}}, \bibinfo {author} {\bibfnamefont {Jos{\'{e}} R.~Crespo}\ \bibnamefont {L{\'{o}}pez-Urrutia}}, \bibinfo {author} {\bibfnamefont {Keisuke}\ \bibnamefont {Fujii}}, \bibinfo {author} {\bibfnamefont {Ren{\'{e}}}\ \bibnamefont {Steinbr{\"{u}}gge}}, \bibinfo {author} {\bibfnamefont {Jakob}\ \bibnamefont {Stierhof}}, \bibinfo {author} {\bibfnamefont {Moto}\ \bibnamefont {Togawa}}, \bibinfo {author} {\bibfnamefont {Zolt{\'{a}}n}\ \bibnamefont {Harman}}, \bibinfo {author} {\bibfnamefont {Natalia~S.}\ \bibnamefont {Oreshkina}}, \bibinfo {author} {\bibfnamefont {Charles}\ \bibnamefont {Cheung}}, \bibinfo {author} {\bibfnamefont {Mikhail~G.}\ \bibnamefont {Kozlov}}, \bibinfo {author} {\bibfnamefont {Sergey~G.}\ \bibnamefont {Porsev}}, \bibinfo {author} {\bibfnamefont {Marianna~S.}\ \bibnamefont {Safronova}}, \bibinfo {author} {\bibfnamefont {Julian~C.}\
  \bibnamefont {Berengut}}, \bibinfo {author} {\bibfnamefont {Michael}\ \bibnamefont {Rosner}}, \bibinfo {author} {\bibfnamefont {Matthias}\ \bibnamefont {Bissinger}}, \bibinfo {author} {\bibfnamefont {Ralf}\ \bibnamefont {Ballhausen}}, \bibinfo {author} {\bibfnamefont {Natalie}\ \bibnamefont {Hell}}, \bibinfo {author} {\bibfnamefont {SungNam}\ \bibnamefont {Park}}, \bibinfo {author} {\bibfnamefont {Moses}\ \bibnamefont {Chung}}, \bibinfo {author} {\bibfnamefont {Moritz}\ \bibnamefont {Hoesch}}, \bibinfo {author} {\bibfnamefont {J{\"{o}}rn}\ \bibnamefont {Seltmann}}, \bibinfo {author} {\bibfnamefont {Andrey~S.}\ \bibnamefont {Surzhykov}}, \bibinfo {author} {\bibfnamefont {Vladimir~A.}\ \bibnamefont {Yerokhin}}, \bibinfo {author} {\bibfnamefont {J{\"{o}}rn}\ \bibnamefont {Wilms}}, \bibinfo {author} {\bibfnamefont {F.~Scott}\ \bibnamefont {Porter}}, \bibinfo {author} {\bibfnamefont {Thomas}\ \bibnamefont {St{\"{o}}hlker}}, \bibinfo {author} {\bibfnamefont {Christoph~H.}\ \bibnamefont {Keitel}}, \bibinfo
  {author} {\bibfnamefont {Thomas}\ \bibnamefont {Pfeifer}}, \bibinfo {author} {\bibfnamefont {Gregory~V.}\ \bibnamefont {Brown}}, \bibinfo {author} {\bibfnamefont {Maurice~A.}\ \bibnamefont {Leutenegger}}, \ and\ \bibinfo {author} {\bibfnamefont {Sven}\ \bibnamefont {Bernitt}},\ }\bibfield  {title} {\enquote {\bibinfo {title} {High resolution photoexcitation measurements exacerbate the long-standing {Fe} {XVII} oscillator strength problem},}\ }\href {\doibase 10.1103/physrevlett.124.225001} {\bibfield  {journal} {\bibinfo  {journal} {Physical Review Letters}\ }\textbf {\bibinfo {volume} {124}},\ \bibinfo {pages} {225001} (\bibinfo {year} {2020})},\ \Eprint {http://arxiv.org/abs/1911.09707} {arXiv:1911.09707} \BibitemShut {NoStop}%
\bibitem [{\citenamefont {Kühn}\ \emph {et~al.}(2022)\citenamefont {Kühn}, \citenamefont {Cheung}, \citenamefont {Oreshkina}, \citenamefont {Steinbrügge}, \citenamefont {Togawa}, \citenamefont {Bernitt}, \citenamefont {Berger}, \citenamefont {Buck}, \citenamefont {Hoesch}, \citenamefont {Seltmann}, \citenamefont {Trinter}, \citenamefont {Keitel}, \citenamefont {Kozlov}, \citenamefont {Porsev}, \citenamefont {Gu}, \citenamefont {Porter}, \citenamefont {Pfeifer}, \citenamefont {Leutenegger}, \citenamefont {Harman}, \citenamefont {Safronova}, \citenamefont {L{\'{o}}pez-Urrutia},\ and\ \citenamefont {Shah}}]{KCO22}%
  \BibitemOpen
  \bibfield  {author} {\bibinfo {author} {\bibfnamefont {S.}~\bibnamefont {Kühn}}, \bibinfo {author} {\bibfnamefont {C.}~\bibnamefont {Cheung}}, \bibinfo {author} {\bibfnamefont {N.~S.}\ \bibnamefont {Oreshkina}}, \bibinfo {author} {\bibfnamefont {Ren{\'{e}}}\ \bibnamefont {Steinbrügge}}, \bibinfo {author} {\bibfnamefont {M.}~\bibnamefont {Togawa}}, \bibinfo {author} {\bibfnamefont {S.}~\bibnamefont {Bernitt}}, \bibinfo {author} {\bibfnamefont {L.}~\bibnamefont {Berger}}, \bibinfo {author} {\bibfnamefont {J.}~\bibnamefont {Buck}}, \bibinfo {author} {\bibfnamefont {M.}~\bibnamefont {Hoesch}}, \bibinfo {author} {\bibfnamefont {J.}~\bibnamefont {Seltmann}}, \bibinfo {author} {\bibfnamefont {F.}~\bibnamefont {Trinter}}, \bibinfo {author} {\bibfnamefont {C.~H.}\ \bibnamefont {Keitel}}, \bibinfo {author} {\bibfnamefont {M.~G.}\ \bibnamefont {Kozlov}}, \bibinfo {author} {\bibfnamefont {S.~G.}\ \bibnamefont {Porsev}}, \bibinfo {author} {\bibfnamefont {M.~F.}\ \bibnamefont {Gu}}, \bibinfo {author} {\bibfnamefont
  {F.~S.}\ \bibnamefont {Porter}}, \bibinfo {author} {\bibfnamefont {T.}~\bibnamefont {Pfeifer}}, \bibinfo {author} {\bibfnamefont {M.~A.}\ \bibnamefont {Leutenegger}}, \bibinfo {author} {\bibfnamefont {Z.}~\bibnamefont {Harman}}, \bibinfo {author} {\bibfnamefont {M.~S.}\ \bibnamefont {Safronova}}, \bibinfo {author} {\bibfnamefont {J.~R.~C.}\ \bibnamefont {L{\'{o}}pez-Urrutia}}, \ and\ \bibinfo {author} {\bibfnamefont {C.}~\bibnamefont {Shah}},\ }\bibfield  {title} {\enquote {\bibinfo {title} {{New Measurement Resolves Key Astrophysical Fe {XVII} Oscillator Strength Problem}},}\ }\href {\doibase 10.1103/physrevlett.129.245001} {\bibfield  {journal} {\bibinfo  {journal} {Physical Review Letters}\ }\textbf {\bibinfo {volume} {129}},\ \bibinfo {pages} {245001} (\bibinfo {year} {2022})},\ \Eprint {http://arxiv.org/abs/2201.09070} {arXiv:2201.09070} \BibitemShut {NoStop}%
\bibitem [{\citenamefont {Shah}\ \emph {et~al.}(2024)\citenamefont {Shah}, \citenamefont {Kühn}, \citenamefont {Bernitt}, \citenamefont {Steinbrügge}, \citenamefont {Togawa}, \citenamefont {Berger}, \citenamefont {Buck}, \citenamefont {Hoesch}, \citenamefont {Seltmann}, \citenamefont {Kozlov}, \citenamefont {Porsev}, \citenamefont {Gu}, \citenamefont {{Scott Porter}}, \citenamefont {Pfeifer}, \citenamefont {Leutenegger}, \citenamefont {Cheung}, \citenamefont {Safronova},\ and\ \citenamefont {L{\'{o}}pez-Urrutia}}]{SKBS24}%
  \BibitemOpen
  \bibfield  {author} {\bibinfo {author} {\bibfnamefont {Chintan}\ \bibnamefont {Shah}}, \bibinfo {author} {\bibfnamefont {Steffen}\ \bibnamefont {Kühn}}, \bibinfo {author} {\bibfnamefont {Sonja}\ \bibnamefont {Bernitt}}, \bibinfo {author} {\bibfnamefont {Ren{\'{e}}}\ \bibnamefont {Steinbrügge}}, \bibinfo {author} {\bibfnamefont {Moto}\ \bibnamefont {Togawa}}, \bibinfo {author} {\bibfnamefont {Lukas}\ \bibnamefont {Berger}}, \bibinfo {author} {\bibfnamefont {Jens}\ \bibnamefont {Buck}}, \bibinfo {author} {\bibfnamefont {Moritz}\ \bibnamefont {Hoesch}}, \bibinfo {author} {\bibfnamefont {Jörn}\ \bibnamefont {Seltmann}}, \bibinfo {author} {\bibfnamefont {Mikhail~G.}\ \bibnamefont {Kozlov}}, \bibinfo {author} {\bibfnamefont {Sergey~G.}\ \bibnamefont {Porsev}}, \bibinfo {author} {\bibfnamefont {Ming~Feng}\ \bibnamefont {Gu}}, \bibinfo {author} {\bibfnamefont {F.}~\bibnamefont {{Scott Porter}}}, \bibinfo {author} {\bibfnamefont {Thomas}\ \bibnamefont {Pfeifer}}, \bibinfo {author} {\bibfnamefont {Maurice~A.}\
  \bibnamefont {Leutenegger}}, \bibinfo {author} {\bibfnamefont {Charles}\ \bibnamefont {Cheung}}, \bibinfo {author} {\bibfnamefont {Marianna~S.}\ \bibnamefont {Safronova}}, \ and\ \bibinfo {author} {\bibfnamefont {Jos{\'{e}} R.~Crespo}\ \bibnamefont {L{\'{o}}pez-Urrutia}},\ }\bibfield  {title} {\enquote {\bibinfo {title} {{Natural-linewidth measurements of the 3C and 3D soft-x-ray transitions in Ni XIX}},}\ }\href {\doibase 10.1103/PhysRevA.109.063108} {\bibfield  {journal} {\bibinfo  {journal} {Phys. Rev. A}\ }\textbf {\bibinfo {volume} {109}},\ \bibinfo {pages} {063108} (\bibinfo {year} {2024})}\BibitemShut {NoStop}%
\bibitem [{\citenamefont {{Sapirstein}}\ \emph {et~al.}(2004)\citenamefont {{Sapirstein}}, \citenamefont {{Pachucki}},\ and\ \citenamefont {{Cheng}}}]{SaPaChe04}%
  \BibitemOpen
  \bibfield  {author} {\bibinfo {author} {\bibfnamefont {J.}~\bibnamefont {{Sapirstein}}}, \bibinfo {author} {\bibfnamefont {K.}~\bibnamefont {{Pachucki}}}, \ and\ \bibinfo {author} {\bibfnamefont {K.~T.}\ \bibnamefont {{Cheng}}},\ }\bibfield  {title} {\enquote {\bibinfo {title} {{Radiative corrections to one-photon decays of hydrogenic ions}},}\ }\href {\doibase 10.1103/PhysRevA.69.022113} {\bibfield  {journal} {\bibinfo  {journal} {Phys. Rev. A}\ }\textbf {\bibinfo {volume} {69}},\ \bibinfo {eid} {022113} (\bibinfo {year} {2004})},\ \Eprint {http://arxiv.org/abs/hep-ph/0311134} {arXiv:hep-ph/0311134} \BibitemShut {NoStop}%
\bibitem [{\citenamefont {Indelicato}\ \emph {et~al.}(2004)\citenamefont {Indelicato}, \citenamefont {Shabaev},\ and\ \citenamefont {Volotka}}]{Indelicato2004}%
  \BibitemOpen
  \bibfield  {author} {\bibinfo {author} {\bibfnamefont {P.}~\bibnamefont {Indelicato}}, \bibinfo {author} {\bibfnamefont {V.~M.}\ \bibnamefont {Shabaev}}, \ and\ \bibinfo {author} {\bibfnamefont {A.~V.}\ \bibnamefont {Volotka}},\ }\bibfield  {title} {\enquote {\bibinfo {title} {{Interelectronic-interaction effect on the transition probability in high-$Z$ He-like ions}},}\ }\href {\doibase 10.1103/physreva.69.062506} {\bibfield  {journal} {\bibinfo  {journal} {Physical Review A}\ }\textbf {\bibinfo {volume} {69}},\ \bibinfo {pages} {062506} (\bibinfo {year} {2004})}\BibitemShut {NoStop}%
\bibitem [{\citenamefont {Andreev}\ \emph {et~al.}(2009)\citenamefont {Andreev}, \citenamefont {Labzowsky},\ and\ \citenamefont {Plunien}}]{Andreev2009}%
  \BibitemOpen
  \bibfield  {author} {\bibinfo {author} {\bibfnamefont {Oleg~Yu.}\ \bibnamefont {Andreev}}, \bibinfo {author} {\bibfnamefont {Leonti~N.}\ \bibnamefont {Labzowsky}}, \ and\ \bibinfo {author} {\bibfnamefont {Günter}\ \bibnamefont {Plunien}},\ }\bibfield  {title} {\enquote {\bibinfo {title} {{QED} calculation of transition probabilities in two-electron ions},}\ }\href {\doibase 10.1103/physreva.79.032515} {\bibfield  {journal} {\bibinfo  {journal} {Physical Review A}\ }\textbf {\bibinfo {volume} {79}},\ \bibinfo {pages} {032515} (\bibinfo {year} {2009})}\BibitemShut {NoStop}%
\bibitem [{\citenamefont {Wood}\ \emph {et~al.}(1997)\citenamefont {Wood}, \citenamefont {Bennett}, \citenamefont {Cho}, \citenamefont {Masterson}, \citenamefont {Roberts}, \citenamefont {Tanner},\ and\ \citenamefont {Wieman}}]{WBC97}%
  \BibitemOpen
  \bibfield  {author} {\bibinfo {author} {\bibfnamefont {C.~S.}\ \bibnamefont {Wood}}, \bibinfo {author} {\bibfnamefont {S.~C.}\ \bibnamefont {Bennett}}, \bibinfo {author} {\bibfnamefont {D.}~\bibnamefont {Cho}}, \bibinfo {author} {\bibfnamefont {B.~P.}\ \bibnamefont {Masterson}}, \bibinfo {author} {\bibfnamefont {J.~L.}\ \bibnamefont {Roberts}}, \bibinfo {author} {\bibfnamefont {C.~E.}\ \bibnamefont {Tanner}}, \ and\ \bibinfo {author} {\bibfnamefont {C.~E.}\ \bibnamefont {Wieman}},\ }\bibfield  {title} {\enquote {\bibinfo {title} {Measurement of parity nonconservation and an anapole moment in cesium},}\ }\href {\doibase 10.1126/science.275.5307.1759} {\bibfield  {journal} {\bibinfo  {journal} {Science}\ }\textbf {\bibinfo {volume} {275}},\ \bibinfo {pages} {1759} (\bibinfo {year} {1997})}\BibitemShut {NoStop}%
\bibitem [{\citenamefont {Safronova}\ \emph {et~al.}(2018)\citenamefont {Safronova}, \citenamefont {Budker}, \citenamefont {DeMille}, \citenamefont {{Jackson Kimball}}, \citenamefont {Derevianko},\ and\ \citenamefont {Clark}}]{SBDJ18}%
  \BibitemOpen
  \bibfield  {author} {\bibinfo {author} {\bibfnamefont {M.~S.}\ \bibnamefont {Safronova}}, \bibinfo {author} {\bibfnamefont {D.}~\bibnamefont {Budker}}, \bibinfo {author} {\bibfnamefont {D.}~\bibnamefont {DeMille}}, \bibinfo {author} {\bibfnamefont {D.~F.}\ \bibnamefont {{Jackson Kimball}}}, \bibinfo {author} {\bibfnamefont {A.}~\bibnamefont {Derevianko}}, \ and\ \bibinfo {author} {\bibfnamefont {C.~W.}\ \bibnamefont {Clark}},\ }\bibfield  {title} {\enquote {\bibinfo {title} {{Search for New Physics with Atoms and Molecules}},}\ }\href@noop {} {\bibfield  {journal} {\bibinfo  {journal} {Rev. Mod. Phys.}\ }\textbf {\bibinfo {volume} {90}},\ \bibinfo {eid} {025008} (\bibinfo {year} {2018})},\ \Eprint {http://arxiv.org/abs/1710.01833} {arXiv:1710.01833} \BibitemShut {NoStop}%
\bibitem [{\citenamefont {Sapirstein}\ and\ \citenamefont {Cheng}(2005)}]{SaChe05}%
  \BibitemOpen
  \bibfield  {author} {\bibinfo {author} {\bibfnamefont {J.}~\bibnamefont {Sapirstein}}\ and\ \bibinfo {author} {\bibfnamefont {K.~T.}\ \bibnamefont {Cheng}},\ }\bibfield  {title} {\enquote {\bibinfo {title} {{Calculation of radiative corrections to $E1$ matrix elements in the neutral alkali metals}},}\ }\href {\doibase 10.1103/PhysRevA.71.022503} {\bibfield  {journal} {\bibinfo  {journal} {Phys. Rev. A}\ }\textbf {\bibinfo {volume} {71}},\ \bibinfo {pages} {022503} (\bibinfo {year} {2005})}\BibitemShut {NoStop}%
\bibitem [{\citenamefont {Fairhall}\ \emph {et~al.}(2023)\citenamefont {Fairhall}, \citenamefont {Roberts},\ and\ \citenamefont {Ginges}}]{Fairhall2023}%
  \BibitemOpen
  \bibfield  {author} {\bibinfo {author} {\bibfnamefont {C.~J.}\ \bibnamefont {Fairhall}}, \bibinfo {author} {\bibfnamefont {B.~M.}\ \bibnamefont {Roberts}}, \ and\ \bibinfo {author} {\bibfnamefont {J.~S.~M.}\ \bibnamefont {Ginges}},\ }\bibfield  {title} {\enquote {\bibinfo {title} {{QED} radiative corrections to electric dipole amplitudes in heavy atoms},}\ }\href {\doibase 10.1103/physreva.107.022813} {\bibfield  {journal} {\bibinfo  {journal} {Physical Review A}\ }\textbf {\bibinfo {volume} {107}},\ \bibinfo {pages} {022813} (\bibinfo {year} {2023})},\ \Eprint {http://arxiv.org/abs/2212.11490} {arXiv:2212.11490} \BibitemShut {NoStop}%
\bibitem [{\citenamefont {Tran~Tan}\ and\ \citenamefont {Derevianko}(2023)}]{TaDe23}%
  \BibitemOpen
  \bibfield  {author} {\bibinfo {author} {\bibfnamefont {H.~B.}\ \bibnamefont {Tran~Tan}}\ and\ \bibinfo {author} {\bibfnamefont {A.}~\bibnamefont {Derevianko}},\ }\bibfield  {title} {\enquote {\bibinfo {title} {Precision theoretical determination of electric-dipole matrix elements in atomic cesium},}\ }\href {\doibase 10.1103/PhysRevA.107.042809} {\bibfield  {journal} {\bibinfo  {journal} {Phys. Rev. A}\ }\textbf {\bibinfo {volume} {107}},\ \bibinfo {pages} {042809} (\bibinfo {year} {2023})}\BibitemShut {NoStop}%
\bibitem [{\citenamefont {Roberts}\ \emph {et~al.}(2023)\citenamefont {Roberts}, \citenamefont {Fairhall},\ and\ \citenamefont {Ginges}}]{Roberts2023}%
  \BibitemOpen
  \bibfield  {author} {\bibinfo {author} {\bibfnamefont {B.~M.}\ \bibnamefont {Roberts}}, \bibinfo {author} {\bibfnamefont {C.~J.}\ \bibnamefont {Fairhall}}, \ and\ \bibinfo {author} {\bibfnamefont {J.~S.~M.}\ \bibnamefont {Ginges}},\ }\bibfield  {title} {\enquote {\bibinfo {title} {Electric-dipole transition amplitudes for atoms and ions with one valence electron},}\ }\href {\doibase 10.1103/physreva.107.052812} {\bibfield  {journal} {\bibinfo  {journal} {Physical Review A}\ }\textbf {\bibinfo {volume} {107}},\ \bibinfo {pages} {052812} (\bibinfo {year} {2023})},\ \Eprint {http://arxiv.org/abs/2211.11134} {arXiv:2211.11134} \BibitemShut {NoStop}%
\bibitem [{\citenamefont {{Flambaum}}\ and\ \citenamefont {{Ginges}}(2005)}]{FlaGin05}%
  \BibitemOpen
  \bibfield  {author} {\bibinfo {author} {\bibfnamefont {V.~V.}\ \bibnamefont {{Flambaum}}}\ and\ \bibinfo {author} {\bibfnamefont {J.~S.~M.}\ \bibnamefont {{Ginges}}},\ }\bibfield  {title} {\enquote {\bibinfo {title} {{Radiative potential and calculations of QED radiative corrections to energy levels and electromagnetic amplitudes in many-electron atoms}},}\ }\href {\doibase 10.1103/PhysRevA.72.052115} {\bibfield  {journal} {\bibinfo  {journal} {Phys. Rev. A}\ }\textbf {\bibinfo {volume} {72}},\ \bibinfo {pages} {052115} (\bibinfo {year} {2005})}\BibitemShut {NoStop}%
\bibitem [{\citenamefont {Shabaev}(2002)}]{shabaev:02:rep}%
  \BibitemOpen
  \bibfield  {author} {\bibinfo {author} {\bibfnamefont {V.~M.}\ \bibnamefont {Shabaev}},\ }\bibfield  {title} {\enquote {\bibinfo {title} {Two-time {G}reen's function method in quantum electrodynamics of high-{$Z$} few-electron atoms},}\ }\href@noop {} {\bibfield  {journal} {\bibinfo  {journal} {Phys. Rep.}\ }\textbf {\bibinfo {volume} {356}},\ \bibinfo {pages} {119 -- 228} (\bibinfo {year} {2002})}\BibitemShut {NoStop}%
\bibitem [{\citenamefont {Yerokhin}\ and\ \citenamefont {Shabaev}(1999)}]{yerokhin:99:pra}%
  \BibitemOpen
  \bibfield  {author} {\bibinfo {author} {\bibfnamefont {V.~A.}\ \bibnamefont {Yerokhin}}\ and\ \bibinfo {author} {\bibfnamefont {V.~M.}\ \bibnamefont {Shabaev}},\ }\bibfield  {title} {\enquote {\bibinfo {title} {First order self-energy correction in hydrogen-like systems},}\ }\href@noop {} {\bibfield  {journal} {\bibinfo  {journal} {Phys. Rev. A}\ }\textbf {\bibinfo {volume} {60}},\ \bibinfo {pages} {800 -- 811} (\bibinfo {year} {1999})}\BibitemShut {NoStop}%
\bibitem [{\citenamefont {Kohn}\ and\ \citenamefont {Sham}(1965)}]{1965_KohnW_PR140}%
  \BibitemOpen
  \bibfield  {author} {\bibinfo {author} {\bibfnamefont {W.}~\bibnamefont {Kohn}}\ and\ \bibinfo {author} {\bibfnamefont {L.~J.}\ \bibnamefont {Sham}},\ }\bibfield  {title} {\enquote {\bibinfo {title} {Self-{{Consistent Equations Including Exchange}} and {{Correlation Effects}}},}\ }\href {\doibase 10.1103/PhysRev.140.A1133} {\bibfield  {journal} {\bibinfo  {journal} {Physical Review}\ }\textbf {\bibinfo {volume} {140}},\ \bibinfo {pages} {A1133--A1138} (\bibinfo {year} {1965})}\BibitemShut {NoStop}%
\bibitem [{\citenamefont {Yerokhin}(2011)}]{yerokhin:11:fns}%
  \BibitemOpen
  \bibfield  {author} {\bibinfo {author} {\bibfnamefont {Vladimir~A.}\ \bibnamefont {Yerokhin}},\ }\bibfield  {title} {\enquote {\bibinfo {title} {Nuclear-size correction to the {L}amb shift of one-electron atoms},}\ }\href@noop {} {\bibfield  {journal} {\bibinfo  {journal} {Phys. Rev. A}\ }\textbf {\bibinfo {volume} {83}},\ \bibinfo {pages} {012507--1 -- 012507--10} (\bibinfo {year} {2011})}\BibitemShut {NoStop}%
\bibitem [{\citenamefont {Tupitsyn}\ and\ \citenamefont {Berseneva}(2013)}]{TuBe13}%
  \BibitemOpen
  \bibfield  {author} {\bibinfo {author} {\bibfnamefont {I.~I.}\ \bibnamefont {Tupitsyn}}\ and\ \bibinfo {author} {\bibfnamefont {E.~V.}\ \bibnamefont {Berseneva}},\ }\bibfield  {title} {\enquote {\bibinfo {title} {A single-particle nonlocal potential for taking into account quantum-electrodynamic corrections in calculations of the electronic structure of atoms},}\ }\href@noop {} {\bibfield  {journal} {\bibinfo  {journal} {Opt. Spectrosc.}\ }\textbf {\bibinfo {volume} {114}},\ \bibinfo {pages} {682} (\bibinfo {year} {2013})}\BibitemShut {NoStop}%
\bibitem [{\citenamefont {Shabaev}\ \emph {et~al.}(2013)\citenamefont {Shabaev}, \citenamefont {Tupitsyn},\ and\ \citenamefont {Yerokhin}}]{STY13}%
  \BibitemOpen
  \bibfield  {author} {\bibinfo {author} {\bibfnamefont {V.~M.}\ \bibnamefont {Shabaev}}, \bibinfo {author} {\bibfnamefont {I.~I.}\ \bibnamefont {Tupitsyn}}, \ and\ \bibinfo {author} {\bibfnamefont {V.~A.}\ \bibnamefont {Yerokhin}},\ }\bibfield  {title} {\enquote {\bibinfo {title} {Model operator approach to the lamb shift calculations in relativistic many-electron atoms},}\ }\href {\doibase 10.1103/PhysRevA.88.012513} {\bibfield  {journal} {\bibinfo  {journal} {Phys. Rev. A}\ }\textbf {\bibinfo {volume} {88}},\ \bibinfo {pages} {012513} (\bibinfo {year} {2013})}\BibitemShut {NoStop}%
\bibitem [{\citenamefont {Tupitsyn}\ \emph {et~al.}(2016)\citenamefont {Tupitsyn}, \citenamefont {Kozlov}, \citenamefont {Safronova}, \citenamefont {Shabaev},\ and\ \citenamefont {Dzuba}}]{TKSSD16}%
  \BibitemOpen
  \bibfield  {author} {\bibinfo {author} {\bibfnamefont {I.~I.}\ \bibnamefont {Tupitsyn}}, \bibinfo {author} {\bibfnamefont {M.~G.}\ \bibnamefont {Kozlov}}, \bibinfo {author} {\bibfnamefont {M.~S.}\ \bibnamefont {Safronova}}, \bibinfo {author} {\bibfnamefont {V.~M.}\ \bibnamefont {Shabaev}}, \ and\ \bibinfo {author} {\bibfnamefont {V.~A.}\ \bibnamefont {Dzuba}},\ }\bibfield  {title} {\enquote {\bibinfo {title} {{Quantum Electrodynamical Shifts in Multivalent Heavy Ions}},}\ }\href {\doibase 10.1103/PhysRevLett.117.253001} {\bibfield  {journal} {\bibinfo  {journal} {Phys. Rev. Lett.}\ }\textbf {\bibinfo {volume} {117}},\ \bibinfo {pages} {253001} (\bibinfo {year} {2016})},\ \Eprint {http://arxiv.org/abs/1607.07064} {arXiv:1607.07064} \BibitemShut {NoStop}%
\bibitem [{\citenamefont {{Ginges}}\ and\ \citenamefont {{Berengut}}(2016)}]{GiBe16}%
  \BibitemOpen
  \bibfield  {author} {\bibinfo {author} {\bibfnamefont {J.~S.~M.}\ \bibnamefont {{Ginges}}}\ and\ \bibinfo {author} {\bibfnamefont {J.~C.}\ \bibnamefont {{Berengut}}},\ }\bibfield  {title} {\enquote {\bibinfo {title} {{QED radiative corrections and many-body effects in atoms: the Uehling potential and shifts in alkali metals}},}\ }\href {\doibase 10.1088/0953-4075/49/9/095001} {\bibfield  {journal} {\bibinfo  {journal} {Journal of Physics B}\ }\textbf {\bibinfo {volume} {49}},\ \bibinfo {pages} {095001} (\bibinfo {year} {2016})},\ \Eprint {http://arxiv.org/abs/1511.01459} {arXiv:1511.01459} \BibitemShut {NoStop}%
\bibitem [{\citenamefont {Guo}\ \emph {et~al.}(2024)\citenamefont {Guo}, \citenamefont {Pa\v{s}teka}, \citenamefont {Nagame}, \citenamefont {Sato}, \citenamefont {Eliav}, \citenamefont {Reitsma},\ and\ \citenamefont {Borschevsky}}]{GuPaNa24}%
  \BibitemOpen
  \bibfield  {author} {\bibinfo {author} {\bibfnamefont {Yangyang}\ \bibnamefont {Guo}}, \bibinfo {author} {\bibfnamefont {Luk\'a\v{s}~F.}\ \bibnamefont {Pa\v{s}teka}}, \bibinfo {author} {\bibfnamefont {Yuichiro}\ \bibnamefont {Nagame}}, \bibinfo {author} {\bibfnamefont {Tetsuya~K.}\ \bibnamefont {Sato}}, \bibinfo {author} {\bibfnamefont {Ephraim}\ \bibnamefont {Eliav}}, \bibinfo {author} {\bibfnamefont {Marten~L.}\ \bibnamefont {Reitsma}}, \ and\ \bibinfo {author} {\bibfnamefont {Anastasia}\ \bibnamefont {Borschevsky}},\ }\bibfield  {title} {\enquote {\bibinfo {title} {Relativistic coupled-cluster calculations of the electron affinity and ionization potentials of lawrencium},}\ }\href {\doibase 10.1103/PhysRevA.110.022817} {\bibfield  {journal} {\bibinfo  {journal} {Phys. Rev. A}\ }\textbf {\bibinfo {volume} {110}},\ \bibinfo {pages} {022817} (\bibinfo {year} {2024})}\BibitemShut {NoStop}%
\bibitem [{\citenamefont {Skripnikov}\ \emph {et~al.}(2024)\citenamefont {Skripnikov}, \citenamefont {Prosnyak}, \citenamefont {Malyshev}, \citenamefont {Athanasakis-Kaklamanakis}, \citenamefont {Brinson}, \citenamefont {Minamisono}, \citenamefont {Cruz}, \citenamefont {Reilly}, \citenamefont {Rickey},\ and\ \citenamefont {Ruiz}}]{SkPrMa24}%
  \BibitemOpen
  \bibfield  {author} {\bibinfo {author} {\bibfnamefont {Leonid~V.}\ \bibnamefont {Skripnikov}}, \bibinfo {author} {\bibfnamefont {Sergey~D.}\ \bibnamefont {Prosnyak}}, \bibinfo {author} {\bibfnamefont {Aleksei~V.}\ \bibnamefont {Malyshev}}, \bibinfo {author} {\bibfnamefont {Michail}\ \bibnamefont {Athanasakis-Kaklamanakis}}, \bibinfo {author} {\bibfnamefont {Alex~Jose}\ \bibnamefont {Brinson}}, \bibinfo {author} {\bibfnamefont {Kei}\ \bibnamefont {Minamisono}}, \bibinfo {author} {\bibfnamefont {Fabian C.~Pastrana}\ \bibnamefont {Cruz}}, \bibinfo {author} {\bibfnamefont {Jordan~Ray}\ \bibnamefont {Reilly}}, \bibinfo {author} {\bibfnamefont {Brooke~J.}\ \bibnamefont {Rickey}}, \ and\ \bibinfo {author} {\bibfnamefont {Ronald. F.~Garcia}\ \bibnamefont {Ruiz}},\ }\bibfield  {title} {\enquote {\bibinfo {title} {Isotope-shift factors with quantum electrodynamics effects for many-electron systems: A study of the nuclear charge radius of $^{26m}\mathrm{Al}$},}\ }\href {\doibase 10.1103/PhysRevA.110.012807} {\bibfield
   {journal} {\bibinfo  {journal} {Phys. Rev. A}\ }\textbf {\bibinfo {volume} {110}},\ \bibinfo {pages} {012807} (\bibinfo {year} {2024})}\BibitemShut {NoStop}%
\bibitem [{\citenamefont {Porsev}\ \emph {et~al.}(2020)\citenamefont {Porsev}, \citenamefont {Safronova}, \citenamefont {Safronova}, \citenamefont {Schmidt}, \citenamefont {Bondarev}, \citenamefont {Kozlov}, \citenamefont {Tupitsyn},\ and\ \citenamefont {Cheung}}]{PSSS20}%
  \BibitemOpen
  \bibfield  {author} {\bibinfo {author} {\bibfnamefont {S.~G.}\ \bibnamefont {Porsev}}, \bibinfo {author} {\bibfnamefont {U.~I.}\ \bibnamefont {Safronova}}, \bibinfo {author} {\bibfnamefont {M.~S.}\ \bibnamefont {Safronova}}, \bibinfo {author} {\bibfnamefont {P.~O.}\ \bibnamefont {Schmidt}}, \bibinfo {author} {\bibfnamefont {A.~I.}\ \bibnamefont {Bondarev}}, \bibinfo {author} {\bibfnamefont {M.~G.}\ \bibnamefont {Kozlov}}, \bibinfo {author} {\bibfnamefont {I.~I.}\ \bibnamefont {Tupitsyn}}, \ and\ \bibinfo {author} {\bibfnamefont {C.}~\bibnamefont {Cheung}},\ }\bibfield  {title} {\enquote {\bibinfo {title} {{Optical clocks based on the Cf$^{+15}$ and Cf$^{+17}$ ions}},}\ }\href {\doibase 10.1103/physreva.102.012802} {\bibfield  {journal} {\bibinfo  {journal} {Physical Review A}\ }\textbf {\bibinfo {volume} {102}},\ \bibinfo {pages} {012802} (\bibinfo {year} {2020})},\ \Eprint {http://arxiv.org/abs/2004.05978} {arXiv:2004.05978} \BibitemShut {NoStop}%
\bibitem [{\citenamefont {{Allehabi}}\ \emph {et~al.}(2024)\citenamefont {{Allehabi}}, \citenamefont {{Dzuba}},\ and\ \citenamefont {{Flambaum}}}]{AlDzFl24}%
  \BibitemOpen
  \bibfield  {author} {\bibinfo {author} {\bibfnamefont {Saleh~O.}\ \bibnamefont {{Allehabi}}}, \bibinfo {author} {\bibfnamefont {V.~A.}\ \bibnamefont {{Dzuba}}}, \ and\ \bibinfo {author} {\bibfnamefont {V.~V.}\ \bibnamefont {{Flambaum}}},\ }\bibfield  {title} {\enquote {\bibinfo {title} {{High-accuracy optical clocks with sensitivity to the fine-structure constant variation based on {Sm$^{10+}$}}},}\ }\href {\doibase 10.1016/j.jqsrt.2024.109151} {\bibfield  {journal} {\bibinfo  {journal} {J. Quant. Spectr. Rad. Transfer}\ }\textbf {\bibinfo {volume} {328}},\ \bibinfo {eid} {109151} (\bibinfo {year} {2024})},\ \Eprint {http://arxiv.org/abs/2404.00854} {arXiv:2404.00854 [hep-ph]} \BibitemShut {NoStop}%
\bibitem [{\citenamefont {Skripnikov}\ \emph {et~al.}(2021)\citenamefont {Skripnikov}, \citenamefont {Chubukov},\ and\ \citenamefont {Shakhova}}]{SkChSh21}%
  \BibitemOpen
  \bibfield  {author} {\bibinfo {author} {\bibfnamefont {Leonid~V.}\ \bibnamefont {Skripnikov}}, \bibinfo {author} {\bibfnamefont {Dmitry~V.}\ \bibnamefont {Chubukov}}, \ and\ \bibinfo {author} {\bibfnamefont {Vera~M.}\ \bibnamefont {Shakhova}},\ }\bibfield  {title} {\enquote {\bibinfo {title} {{The role of QED effects in transition energies of heavy-atom alkaline earth monofluoride molecules: A theoretical study of Ba$^+$, BaF, RaF, and E120F}},}\ }\href {\doibase 10.1063/5.0068267} {\bibfield  {journal} {\bibinfo  {journal} {The Journal of Chemical Physics}\ }\textbf {\bibinfo {volume} {155}},\ \bibinfo {pages} {144103} (\bibinfo {year} {2021})}\BibitemShut {NoStop}%
\bibitem [{\citenamefont {Skripnikov}(2021)}]{Sk21}%
  \BibitemOpen
  \bibfield  {author} {\bibinfo {author} {\bibfnamefont {Leonid~V.}\ \bibnamefont {Skripnikov}},\ }\bibfield  {title} {\enquote {\bibinfo {title} {{Approaching meV level for transition energies in the radium monofluoride molecule RaF and radium cation Ra+ by including quantum-electrodynamics effects}},}\ }\href {\doibase 10.1063/5.0053659} {\bibfield  {journal} {\bibinfo  {journal} {The Journal of Chemical Physics}\ }\textbf {\bibinfo {volume} {154}},\ \bibinfo {pages} {201101} (\bibinfo {year} {2021})}\BibitemShut {NoStop}%
\bibitem [{\citenamefont {Wilkins}\ \emph {et~al.}(2024)\citenamefont {Wilkins}, \citenamefont {Perrett}, \citenamefont {Udrescu} \emph {et~al.}}]{WiPeUd24}%
  \BibitemOpen
  \bibfield  {author} {\bibinfo {author} {\bibfnamefont {S.~G.}\ \bibnamefont {Wilkins}}, \bibinfo {author} {\bibfnamefont {H.~A.}\ \bibnamefont {Perrett}}, \bibinfo {author} {\bibfnamefont {S.~M.}\ \bibnamefont {Udrescu}},  \emph {et~al.},\ }\href {https://arxiv.org/abs/2408.14673} {\enquote {\bibinfo {title} {Ionization potential of radium monofluoride},}\ } (\bibinfo {year} {2024}),\ \Eprint {http://arxiv.org/abs/2408.14673} {arXiv:2408.14673 [physics.atom-ph]} \BibitemShut {NoStop}%
\bibitem [{\citenamefont {Eliav}\ \emph {et~al.}(2015)\citenamefont {Eliav}, \citenamefont {Fritzsche},\ and\ \citenamefont {Kaldor}}]{ElFrKa15}%
  \BibitemOpen
  \bibfield  {author} {\bibinfo {author} {\bibfnamefont {Ephraim}\ \bibnamefont {Eliav}}, \bibinfo {author} {\bibfnamefont {Stephan}\ \bibnamefont {Fritzsche}}, \ and\ \bibinfo {author} {\bibfnamefont {Uzi}\ \bibnamefont {Kaldor}},\ }\bibfield  {title} {\enquote {\bibinfo {title} {Electronic structure theory of the superheavy elements},}\ }\href {\doibase 10.1016/j.nuclphysa.2015.06.017} {\bibfield  {journal} {\bibinfo  {journal} {Nuclear Physics A}\ }\textbf {\bibinfo {volume} {944}},\ \bibinfo {pages} {518--550} (\bibinfo {year} {2015})},\ \bibinfo {note} {special Issue on Superheavy Elements}\BibitemShut {NoStop}%
\bibitem [{\citenamefont {Malyshev}\ \emph {et~al.}(2022)\citenamefont {Malyshev}, \citenamefont {Glazov}, \citenamefont {Shabaev}, \citenamefont {Tupitsyn}, \citenamefont {Yerokhin},\ and\ \citenamefont {Zaytsev}}]{MaGlSh22}%
  \BibitemOpen
  \bibfield  {author} {\bibinfo {author} {\bibfnamefont {A.~V.}\ \bibnamefont {Malyshev}}, \bibinfo {author} {\bibfnamefont {D.~A.}\ \bibnamefont {Glazov}}, \bibinfo {author} {\bibfnamefont {V.~M.}\ \bibnamefont {Shabaev}}, \bibinfo {author} {\bibfnamefont {I.~I.}\ \bibnamefont {Tupitsyn}}, \bibinfo {author} {\bibfnamefont {V.~A.}\ \bibnamefont {Yerokhin}}, \ and\ \bibinfo {author} {\bibfnamefont {V.~A.}\ \bibnamefont {Zaytsev}},\ }\bibfield  {title} {\enquote {\bibinfo {title} {{Model-QED operator for superheavy elements}},}\ }\href {\doibase 10.1103/PhysRevA.106.012806} {\bibfield  {journal} {\bibinfo  {journal} {Phys. Rev. A}\ }\textbf {\bibinfo {volume} {106}},\ \bibinfo {pages} {012806} (\bibinfo {year} {2022})}\BibitemShut {NoStop}%
\bibitem [{\citenamefont {Shabaev}\ \emph {et~al.}(2015)\citenamefont {Shabaev}, \citenamefont {Tupitsyn},\ and\ \citenamefont {Yerokhin}}]{STY15}%
  \BibitemOpen
  \bibfield  {author} {\bibinfo {author} {\bibfnamefont {V.~M.}\ \bibnamefont {Shabaev}}, \bibinfo {author} {\bibfnamefont {I.~I.}\ \bibnamefont {Tupitsyn}}, \ and\ \bibinfo {author} {\bibfnamefont {V.~A.}\ \bibnamefont {Yerokhin}},\ }\bibfield  {title} {\enquote {\bibinfo {title} {{QEDMOD: Fortran program for calculating the model Lamb-shift operator}},}\ }\href@noop {} {\bibfield  {journal} {\bibinfo  {journal} {Comp. Phys. Comm.}\ }\textbf {\bibinfo {volume} {189}},\ \bibinfo {pages} {175--181} (\bibinfo {year} {2015})}\BibitemShut {NoStop}%
\bibitem [{\citenamefont {Shabaev}\ \emph {et~al.}(2018)\citenamefont {Shabaev}, \citenamefont {Tupitsyn},\ and\ \citenamefont {Yerokhin}}]{shabaev:18:qedmod}%
  \BibitemOpen
  \bibfield  {author} {\bibinfo {author} {\bibfnamefont {V.~M.}\ \bibnamefont {Shabaev}}, \bibinfo {author} {\bibfnamefont {I.~I.}\ \bibnamefont {Tupitsyn}}, \ and\ \bibinfo {author} {\bibfnamefont {V.~A.}\ \bibnamefont {Yerokhin}},\ }\bibfield  {title} {\enquote {\bibinfo {title} {{QEDMOD: Fortran program for calculating the model Lamb-shift operator}},}\ }\href {\doibase https://doi.org/10.1016/j.cpc.2017.10.007} {\bibfield  {journal} {\bibinfo  {journal} {Comp. Phys. Comm.}\ }\textbf {\bibinfo {volume} {223}},\ \bibinfo {pages} {69} (\bibinfo {year} {2018})}\BibitemShut {NoStop}%
\bibitem [{\citenamefont {Kramida}\ \emph {et~al.}(2016)\citenamefont {Kramida}, \citenamefont {Ralchenko}, \citenamefont {Reader},\ and\ \citenamefont {{NIST ASD Team}}}]{NIST}%
  \BibitemOpen
  \bibfield  {author} {\bibinfo {author} {\bibfnamefont {A.}~\bibnamefont {Kramida}}, \bibinfo {author} {\bibfnamefont {Yu.}\ \bibnamefont {Ralchenko}}, \bibinfo {author} {\bibfnamefont {J.}~\bibnamefont {Reader}}, \ and\ \bibinfo {author} {\bibnamefont {{NIST ASD Team}}},\ }\href {http://physics.nist.gov/PhysRefData/ASD/index.html} {\enquote {\bibinfo {title} {{NIST Atomic Spectra Database}},}\ } (\bibinfo {year} {2016})\BibitemShut {NoStop}%
\bibitem [{\citenamefont {Kozlov}\ \emph {et~al.}(2015)\citenamefont {Kozlov}, \citenamefont {Porsev}, \citenamefont {Safronova},\ and\ \citenamefont {Tupitsyn}}]{KPST15}%
  \BibitemOpen
  \bibfield  {author} {\bibinfo {author} {\bibfnamefont {M.~G.}\ \bibnamefont {Kozlov}}, \bibinfo {author} {\bibfnamefont {S.~G.}\ \bibnamefont {Porsev}}, \bibinfo {author} {\bibfnamefont {M.~S.}\ \bibnamefont {Safronova}}, \ and\ \bibinfo {author} {\bibfnamefont {I.~I.}\ \bibnamefont {Tupitsyn}},\ }\bibfield  {title} {\enquote {\bibinfo {title} {{CI-MBPT: A package of programs for relativistic atomic calculations based on a method combining configuration interaction and many-body perturbation theory}},}\ }\href {\doibase 10.1016/j.cpc.2015.05.007} {\bibfield  {journal} {\bibinfo  {journal} {Comput. Phys. Commun.}\ }\textbf {\bibinfo {volume} {195}},\ \bibinfo {pages} {199--213} (\bibinfo {year} {2015})}\BibitemShut {NoStop}%
\bibitem [{\citenamefont {Kozlov}\ \emph {et~al.}(2024)\citenamefont {Kozlov}, \citenamefont {Demidov}, \citenamefont {Kaygorodov},\ and\ \citenamefont {Tryapitsyna}}]{KoDeKa24}%
  \BibitemOpen
  \bibfield  {author} {\bibinfo {author} {\bibfnamefont {Mikhail~G.}\ \bibnamefont {Kozlov}}, \bibinfo {author} {\bibfnamefont {Yuriy~A.}\ \bibnamefont {Demidov}}, \bibinfo {author} {\bibfnamefont {Mikhail~Y.}\ \bibnamefont {Kaygorodov}}, \ and\ \bibinfo {author} {\bibfnamefont {Elizaveta~V.}\ \bibnamefont {Tryapitsyna}},\ }\bibfield  {title} {\enquote {\bibinfo {title} {Basis set calculations of heavy atoms},}\ }\href {\doibase 10.3390/atoms12010003} {\bibfield  {journal} {\bibinfo  {journal} {Atoms}\ }\textbf {\bibinfo {volume} {12}},\ \bibinfo {pages} {3} (\bibinfo {year} {2024})},\ \Eprint {http://arxiv.org/abs/2312.07782} {arXiv:2312.07782} \BibitemShut {NoStop}%
\bibitem [{\citenamefont {Kozlov}\ and\ \citenamefont {Tupitsyn}(2019)}]{KozTup19}%
  \BibitemOpen
  \bibfield  {author} {\bibinfo {author} {\bibfnamefont {Mikhail}\ \bibnamefont {Kozlov}}\ and\ \bibinfo {author} {\bibfnamefont {Ilya}\ \bibnamefont {Tupitsyn}},\ }\bibfield  {title} {\enquote {\bibinfo {title} {Mixed basis sets for atomic calculations},}\ }\href {\doibase 10.3390/atoms7030092} {\bibfield  {journal} {\bibinfo  {journal} {Atoms}\ }\textbf {\bibinfo {volume} {7}},\ \bibinfo {pages} {92} (\bibinfo {year} {2019})}\BibitemShut {NoStop}%
\bibitem [{\citenamefont {Cheung}\ \emph {et~al.}(2021)\citenamefont {Cheung}, \citenamefont {Safronova},\ and\ \citenamefont {Porsev}}]{Cheung2021}%
  \BibitemOpen
  \bibfield  {author} {\bibinfo {author} {\bibfnamefont {Charles}\ \bibnamefont {Cheung}}, \bibinfo {author} {\bibfnamefont {Marianna}\ \bibnamefont {Safronova}}, \ and\ \bibinfo {author} {\bibfnamefont {Sergey}\ \bibnamefont {Porsev}},\ }\bibfield  {title} {\enquote {\bibinfo {title} {{Scalable Codes for Precision Calculations of Properties of Complex Atomic Systems}},}\ }\href {\doibase 10.3390/sym13040621} {\bibfield  {journal} {\bibinfo  {journal} {Symmetry}\ }\textbf {\bibinfo {volume} {13}},\ \bibinfo {pages} {621} (\bibinfo {year} {2021})}\BibitemShut {NoStop}%
\bibitem [{\citenamefont {Sobelman}(1979)}]{Sob79}%
  \BibitemOpen
  \bibfield  {author} {\bibinfo {author} {\bibfnamefont {I.~I.}\ \bibnamefont {Sobelman}},\ }\href@noop {} {\emph {\bibinfo {title} {Atomic spectra and radiative transitions}}}\ (\bibinfo  {publisher} {Springer-Verlag},\ \bibinfo {address} {Berlin},\ \bibinfo {year} {1979})\BibitemShut {NoStop}%
\end{thebibliography}%
\end{document}